# A Density-Informed Multimodal Artificial Intelligence Framework for Improving Breast Cancer Detection Across All Breast Densities

Siva Teja Kakileti[1], Bharath Govindaraju[1], Sudhakar Sampangi[2], Geetha Manjunath[1]

[1]Niramai Health Analytix Pvt Ltd, Bangalore, India; [2]HCG Cancer Hospital, Bangalore, India

**Abstract**

Mammography, the current standard for breast cancer screening, has reduced sensitivity in women with dense breast tissue, contributing to missed or delayed diagnoses. Thermalytix, an AI-based thermal imaging modality, captures functional vascular and metabolic cues that may complement mammographic structural data. This study investigates whether a breast density–informed multi-modal AI framework can improve cancer detection by dynamically selecting the appropriate imaging modality based on breast tissue composition. A total of 324 women underwent both mammography and thermal imaging. Mammography images were analyzed using a multi-view deep learning model, while Thermalytix assessed thermal images through vascular and thermal radiomics. The proposed framework utilized Mammography AI for fatty breasts and Thermalytix AI for dense breasts, optimizing predictions based on tissue type. This multi-modal AI framework achieved a sensitivity of 94.55% (95% CI: 88.54–100) and specificity of 79.93% (95% CI: 75.14–84.71), outperforming standalone mammography AI (sensitivity 81.82%, specificity 86.25%) and Thermalytix AI (sensitivity 92.73%, specificity 75.46%). Importantly, the sensitivity of Mammography dropped significantly in dense breasts (67.86%) versus fatty breasts (96.30%), whereas Thermalytix AI maintained high and consistent sensitivity in both (92.59% and 92.86%, respectively).This demonstrates that a density-informed multi-modal AI framework can overcome key limitations of unimodal screening and deliver high performance across diverse breast compositions. The proposed framework is interpretable, low-cost, and easily deployable, offering a practical path to improving breast cancer screening outcomes in both high-resource and resource-limited settings.

## 1. Introduction

Breast cancer remains the leading cause of cancer-related mortality among women worldwide, with an estimated 2.3 million new cases and 660,000 deaths in 2022 alone [1,2]. Early detection through systematic screening is central to improving survival outcomes and reducing treatment burden[3–6]. Mammography, the gold standard for breast cancer screening, has been validated in multiple randomized controlled trials for its ability to reduce mortality, particularly in women aged 50 years and older[3–8].

However, mammography exhibits reduced sensitivity in women with dense breast tissue—who comprise approximately 40–50% of women over the age of 40 [9–14].

Dense fibroglandular tissue can obscure malignant lesions, lowering mammographic sensitivity to as little as 47% [12,13]. Compounding this, breast density itself is an independent risk factor for cancer [15]. This dual challenge of under-detection and increased risk has prompted regulatory bodies such as the U.S. Food and Drug Administration to mandate breast density reporting in mammography assessments [16].

In recent years, artificial intelligence (AI) has transformed breast cancer screening of mammography images through advanced computer-aided algorithms. Multiple studies demonstrated that the AI algorithms trained on large mammography datasets have the ability to identify subtle patterns, reduce inter-reader variability, and achieve accuracy comparable to, or in some cases surpassing, that of radiologists [17-19]. These advancements have shown promise in improving cancer detection rates and streamlining workflows in radiology practices. Despite these improvements, the accuracy of mammography AI algorithms on dense breasts is questionable with some studies still reporting lower accuracies on dense breasts [20-21]. This calls for the need of supplemental imaging modalities to aid in detecting breast cancer in women with dense breasts.

Thermalytix, an innovative AI-based breast cancer detection modality, leverages the principles of thermal imaging and advanced algorithms to identify abnormal thermal and vascular patterns indicative of malignancy [22]. Thermal imaging offers several key advantages, including being radiation-free, non-invasive, privacy-aware, and portable. This technology may be particularly suited for women with dense breast tissue, as the higher vascularity and glandular activity in dense breasts may generate more pronounced thermal patterns, potentially aiding in earlier and more accurate cancer detection in this cohort, where traditional imaging techniques may face limitations.

Thermalytix demonstrated promising results in multiple prospective and retrospective studies. Singh et al [23] evaluated Thermalytix on 258 symptomatic women reporting a sensitivity and specificity of 82·5% and 80·5% respectively. In another study, Kakileti et al [24] obtained sensitivity and specificity of 91·02% and 82·39% respectively. The third study by Bansal et al [25] reported an overall sensitivity of Thermalytix of 95·24% and a specificity of 88·58%. Importantly, it has demonstrated high sensitivities>90% even in women with dense breasts [26], where mammography often falls short.

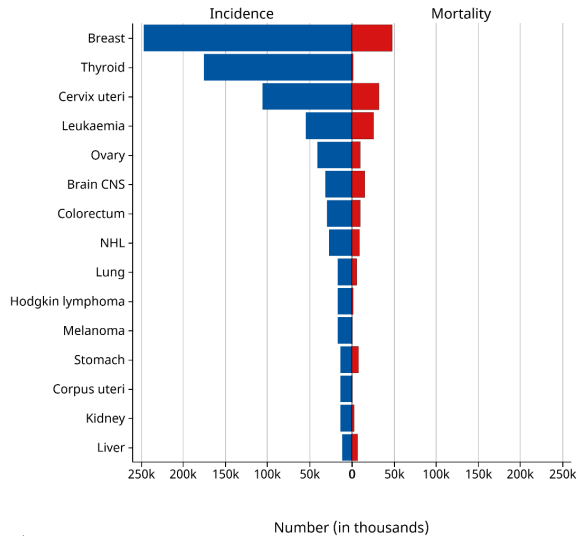

**Figure 1.** Illustration of incidence and mortality rates for different types of cancer worldwide [2].

Given the complementary strengths of mammography and Thermalytix, integrating these modalities within a unified, multi-modal AI framework offers a more comprehensive solution for breast cancer detection. Mammography provides detailed structural imaging, while Thermalytix captures functional cues related to vascular and metabolic activity. By combining these orthogonal insights, the limitations of unimodal screening—particularly in women with dense breasts—may be mitigated. This study proposes and evaluates a breast density–informed multi-modal AI system that dynamically assigns the appropriate imaging modality based on tissue composition, with the aim of improving diagnostic sensitivity in a manner that is clinically scalable, explainable, and equitable.

## 2. Methods

### Study Population

This study utilized anonymized data from women who underwent both Thermalytix and mammography screening at two clinical sites in India (Figure 2). Women aged 18 years and older presenting to the hospital either for routine breast health screening or with clinical symptoms were included. Exclusion criteria encompassed women who were pregnant or lactating, those with a prior history of breast cancer, or those who had undergone previous breast surgery.

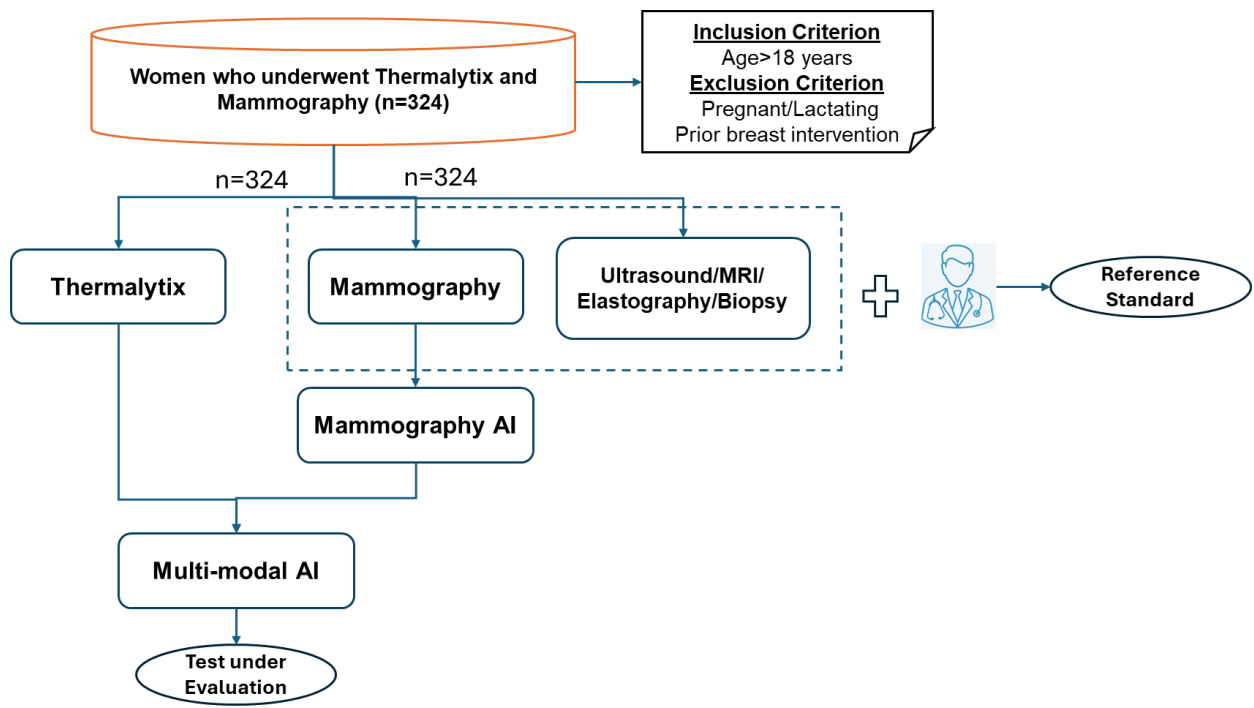

**Figure 2.** Schematic diagram of the study.

A total of 324 women met the inclusion criteria and were enrolled. The median age was 48 years (interquartile range: 43–57 years). The distribution of age, menopausal status, and breast density is shown in Table 1. All participants underwent standard-of-care diagnostic work-up based on institutional protocols, which included mammography and an additional imaging test such as ultrasound or MRI, or histopathological confirmation through biopsy. Prior to these assessments, Thermalytix imaging was performed. Among the 324 women, 55 (17.0%) were classified as suspicious by standard-of-care (SoC) diagnostic work-up.

Informed consent was obtained from all participants prior to their inclusion in the study. Institution Ethics Committee approval from HCG Institute Ethics Committee was obtained to evaluate the multi-modal breast cancer detection with Thermalytix and Mammography. The objective of the study was to evaluate whether a density-informed, multi-modal AI strategy combining mammography and Thermalytix could improve the sensitivity of breast cancer detection relative to either modality alone.

| Variable | Number of women | Number of women classified as suspicious with SoC |
|---|---|---|
| Total | **324 (100%)** | **55 (17.0%)** |
| Age | | |
|     <30 years | 1(0.31%) | 0(0%) |
|     30-39 years | 44(13.58%) | 5(11.3%) |
|     40-49 years | 130(40.12%) | 17(13%) |
|     50-59 years | 87(26.85%) | 16(18.3%) |
|     60-69 years | 51(15.74%) | 11(21.5%) |
|     >=70 years | 11(3.39%) | 6(54.5%) |
| Menopause | | |
|     Yes | 148(45.67%) | 31(20.9%) |
|     No | 176(54.32%) | 24(13.6%) |
| Breast Density | | |
|     ACR category A | 8(2.46%) | 1(12.5%) |
|     ACR category B | 160(49.38%) | 30(18.7%) |
|     ACR category C | 141(43.51%) | 24(17%) |
|     ACR category D | 15(4.62%) | 0(0%) |

Table I. Distribution of study population.

**Mammography AI Interpretation**

A two-stage, multi-view deep learning framework (Figure 3) was developed for automated malignancy detection in mammographic images. The first stage employed a YOLO-X object detection model to crop the breast regions in the standard four-view mammography set: Left Craniocaudal (LCC), Left Mediolateral Oblique (LMLO), Right Craniocaudal (RCC), and Right Mediolateral Oblique (RMLO) [27].

In the second stage, a custom multi-view convolutional neural network (CNN) architecture, with ResNet-22 [28] as the feature extractor, processed each cropped view. Features were concatenated and passed through a dense layer (512 units) followed by a softmax layer to predict malignancy probability

Model training utilized mammographic images from 19,883 subjects compiled from publicly available datasets, including RSNA [29], ViNDR [30], CESM [31], CMMD [32], and Mini-DDSM [33]. A progressive training strategy was adopted, wherein the top layers of the CNN were gradually unfrozen to improve learning efficiency. To improve model robustness and generalizability, a suite of data augmentation techniques was applied during training, including random rotations (within ±10°), horizontal flipping, and the addition of Gaussian noise.

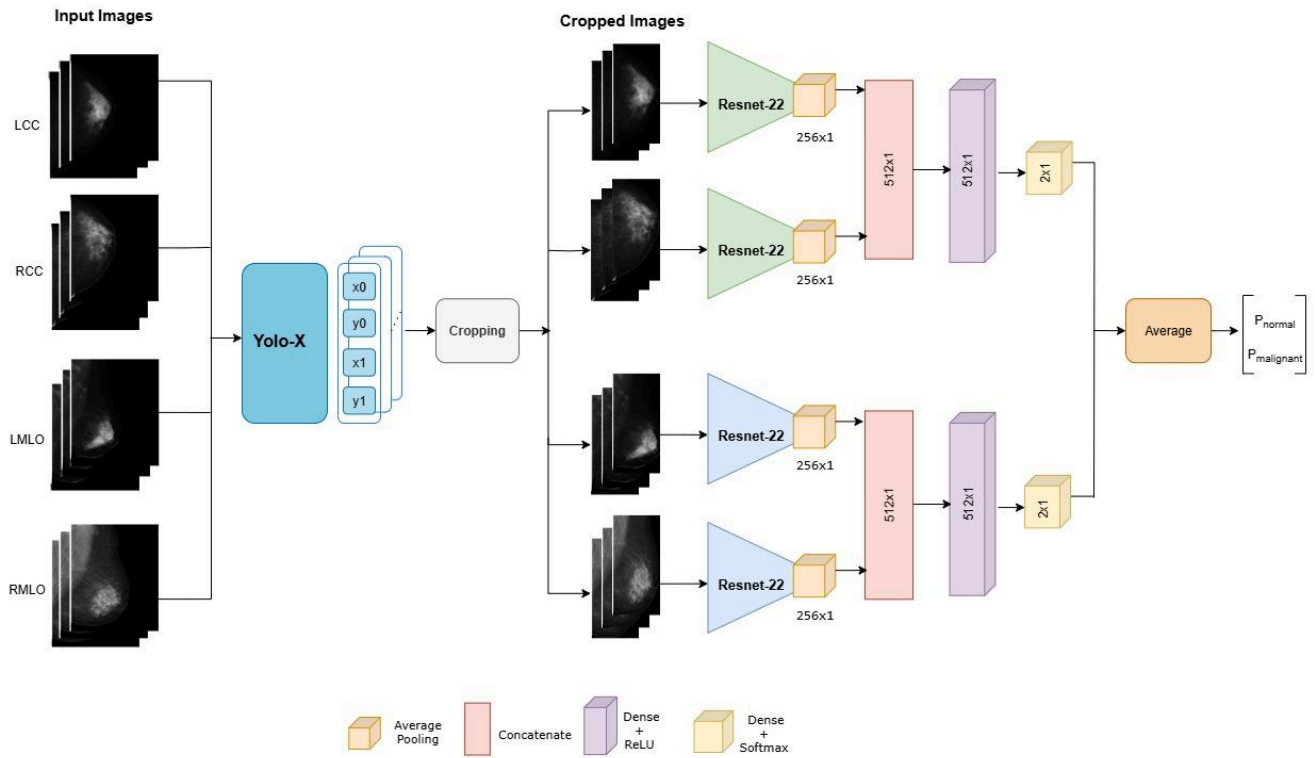

**Figure 3.** Multi-view CNN architecture employed for classification of mammography images.

**Radiomics based Thermalytix AI Interpretation of Thermal Images**

The Thermalytix AI framework integrates deep learning with radiomics-based machine learning algorithms to detect malignancy-associated thermal patterns in an interpretable and explainable manner. Initially, an encoder-decoder convolutional neural network is employed to segment the breast region from the captured thermal images [34-35]. This is followed by three dedicated segmentation modules that delineate hotspots, vascular patterns, and areolar hotspot regions within the segmented breast area, as illustrated in Figure 4 [36-38].

From each of these regions, radiomic features are extracted to quantify thermal asymmetries and vascular morphology [36-38]. Hotspot radiomics capture characteristics such as shape, size, symmetry, and temperature variation of focal thermal activity. Vascular radiomics quantify vessel count, branching complexity, mean vessel caliber, and bilateral symmetry. Areolar radiomics assess thermal asymmetries specifically localized near the areolar region. These handcrafted features encapsulate domain knowledge, enable model interpretability, and contribute to accurate classification performance.

The extracted radiomic features serve as input to three independent machine learning classifiers—each trained to estimate malignancy likelihood from hotspot, vascular, and areolar features, respectively. The resulting malignancy scores are subsequently integrated with clinical metadata in an ensemble model to generate both an ensemble score and the final B-Score. The B-Score, ranging from 1 to 5, stratifies malignancy risk: scores of 1–2 indicate low risk, 3 suggests moderate risk, and 4–5 reflect high malignancy risk. This explainable AI framework fosters clinical trust by providing transparent outputs that support informed diagnostic decision-making. The development and training of the Thermalytix models were based on a large-scale dataset comprising thermal images from over 100,000 women [34–38].

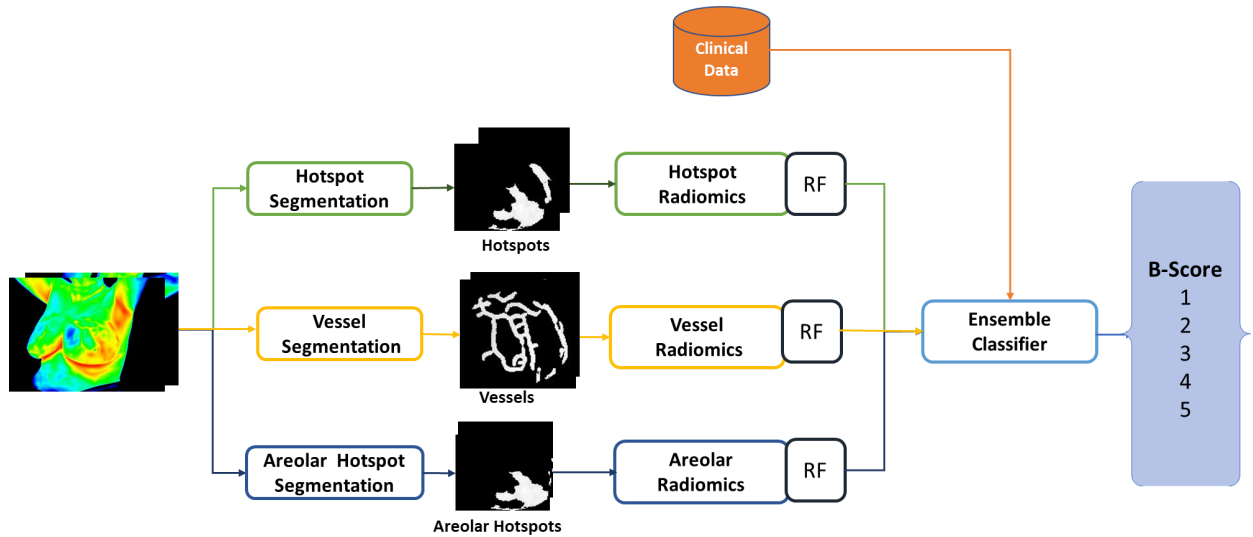

**Figure 4.** Block diagram of Thermalytix AI modules that uses radiomics for classification of thermal images.

**Density-Informed Multimodal AI Framework**

To evaluate the potential of a multi-modal AI-based approach, we implemented a rule-based decision framework guided by breast tissue density. Women with fatty breasts (ACR A/B) were assigned predictions from the mammography AI system, while women with dense breasts (ACR C/D) were assigned predictions from the Thermalytix AI model (Figure 5). This logic was designed to leverage the strengths of each modality, ensuring an optimized screening approach tailored to different tissue types. The output of the mammography AI system was considered positive if the malignancy probability exceeded 0.43, a threshold determined using Youden's Index on a validation dataset. For the Thermalytix AI system, B-scores of 3 or higher were considered test-positive, consistent with prior clinical validations.

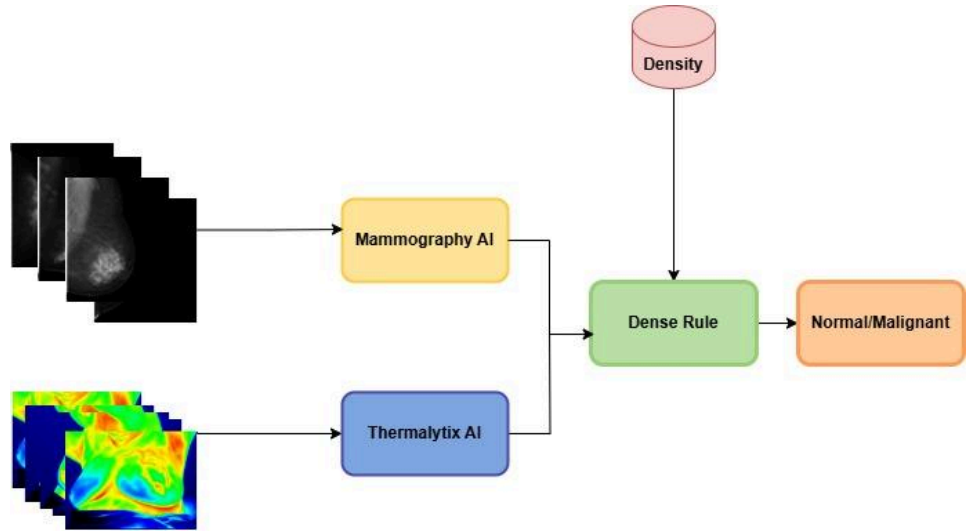

**Figure 5.** Block diagram depicting the Multi-modal AI framework that combines Mammography AI predictions with Thermalytix AI predictions

**Statistical Analysis**

Statistical analysis was conducted using the Python Scikit-Learn and stats frameworks to evaluate the performance of each imaging modality individually and in combination. Key performance metrics included sensitivity, specificity, positive predictive value (PPV), negative predictive value (NPV) and Balanced Accuracy. Confidence intervals were calculated at the 95% level (95CI) using a two-tailed test to ensure statistical reliability of the estimates. The radiologists' conclusions from standard-of-care procedures were used as the reference ground truth for all performance evaluations.

3. Results

A total of 324 women were included in the analysis, with complete mammography and thermal imaging data available. Among these, 168 (51.9%) had fatty breasts (ACR A/B) and 156 (48.1%) had dense breasts (ACR C/D), as determined by radiologist-assigned density scores.

**Performance of Mammography AI algorithm**

Across the full cohort (n = 324), the multi-view mammography AI model achieved a sensitivity of 81.82% (95CI: 71.62%–92.01%), specificity of 86.25% (95% CI: 82.13%–90.36%), positive predictive value (PPV) of 54.88% (95CI: 44.11%–65.65%), and negative predictive value (NPV) of 95.87% (95CI: 93.36%–98.38%).

Among women with fatty breast tissue (n = 168), the model's performance improved substantially, with a sensitivity of 96.30% (95CI: 89.17%–100%), specificity of 89.86% (95% CI: 84.82%–94.89%), PPV of 65.00% (95CI: 50.22%–79.78%), and NPV of 99.20% (95CI: 97.64%–100%).

In contrast, among women with dense breasts (n = 156), sensitivity decreased to 67.86% (95CI: 50.56%–85.16%), specificity to 83.08% (95CI: 76.63%–89.52%), PPV to 46.34% (95CI: 31.08%–61.61%), and NPV to 92.31% (95CI: 87.48%–97.14%).

**Performance of Thermalytix AI algorithm**
Thermalytix demonstrated strong overall sensitivity of 92.73% (95% CI: 85.86%–99.59%), specificity of 75.46% (95% CI: 70.32%–80.61%), PPV of 43.59% (95% CI: 34.40%–52.58%), and NPV of 98.07% (95% CI: 96.19%–99.94%).

Among women with fatty breasts, sensitivity was 92.59% (95% CI: 82.71%–100%), specificity improved to 81.16% (95CI: 74.64%–87.68%), PPV was 49.02% (95% CI: 35.23%–62.74%), and NPV reached 98.25% (95CI: 95.84%–100%).

In dense breasts, sensitivity remained high at 92.86% (95CI: 83.32%–100%), but specificity declined to 69.47% (95CI: 61.58%–77.35%). Corresponding PPV and NPV were 39.39% (95CI: 27.61%–51.12%), and 97.85% (95CI: 94.90%–100%), respectively.

Analysis of the 40 false positives identified by Thermalytix in dense breasts revealed that 31 (77.5%) corresponded to benign but clinically relevant findings, such as fibroadenomas, abscesses, or fibrocystic changes.

**Performance of Density-Informed Multi-modal AI Framework**
In this approach, mammography AI predictions were applied to women with fatty breast tissue, and Thermalytix AI predictions were applied to women with dense breasts. This multi-modal framework achieved a sensitivity of 94.55% (95% CI: 88.54–100), specificity of 79.93% (95% CI: 75.14%–84.71%), PPV of 49.06% (95% CI: 39.54–58.57), and NPV of 98.62% (95% CI: 97.08%–100%). Compared with either modality alone, the multi-modal framework yielded superior sensitivity across the full cohort while preserving clinically acceptable specificity.

By contrast, a naïve OR-rule approach—considering a positive result if either model predicted malignancy—achieved similar sensitivity (94.6%) but reduced specificity (66.5%), underscoring the benefit of informed model selection over undifferentiated combination.

Representative imaging outcomes across both modalities are presented in Figure 7. Summary statistics for all approaches are provided in Table 2, with corresponding ROC curves shown in Figure 6.

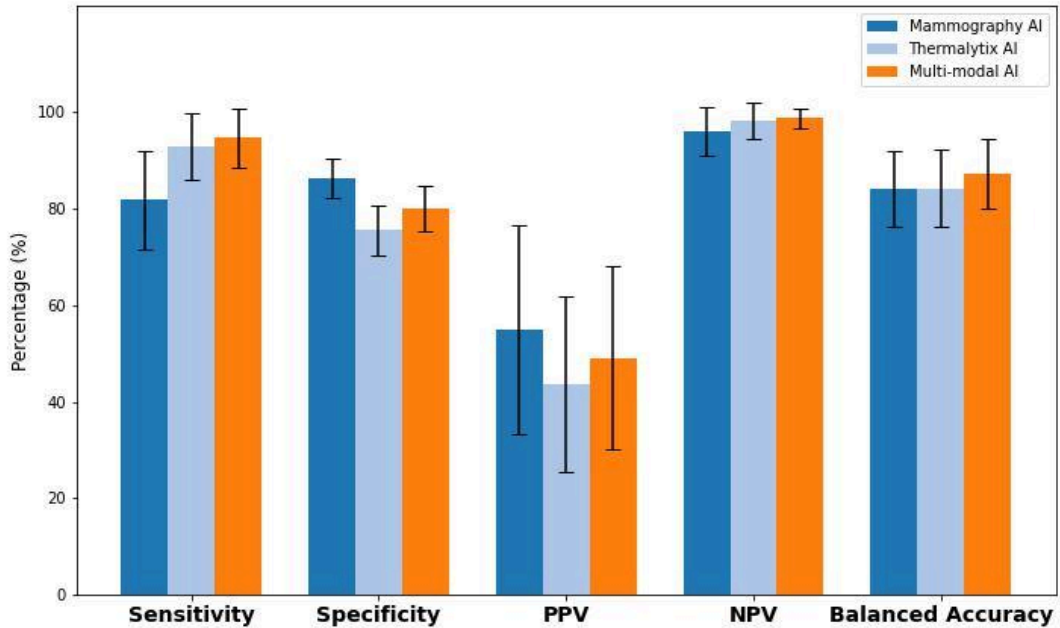

**Figure 6.** Sensitivity, Specificity, PPV and NPV of Mammography AI alone, Thermalytix AI alone and Multi-modal AI.

|  |  | Sensitivity | Specificity | PPV | NPV | Balanced Accuracy |
|---|---|---|---|---|---|---|
| **Mammography AI alone** | Overall | 81.82% (71.6%-92.0%) | 86.25% (82.1%-90.4%) | 54.88% (44.1%-65.7%) | 95.87% (93.4%-98.4%) | 84.04% (80.1%-88.0%) |
|  | Fatty | 96.3% (91.3%-100%) | 89.86% (86.3%-93.5%) | 65.0% (50.2%–79.8%) | 99.20% (97.6%–100%) | 93.08% (90.3%-95.8%) |
|  | Dense | 67.86% (55.5%-80.0%) | 83.08% (78.6%-87.6%) | 46.34% (31.1%–61.6%) | 92.31% (87.5%–97.1%) | 75.47% (70.1%-80.2%) |
| **Thermalytix AI alone** | Overall | 92.73% (85.9%–99.6%) | 75.46% (70.3%–80.6%) | 43.59% (34.4%–52.6%) | 98.07% (96.2%–99.9%) | 84.10% (80.1%-88.1%) |
|  | Fatty | 92.59% (85.7%-99.5%) | 81.16% (76.5%-85.8%) | 49.02% (35.2%–62.7%) | 98.25% (95.8%–100%) | 86.87% (83.2%-90.1%) |
|  | Dense | 92.86% (86.1%-99.7%) | 69.47% (64.0%-75.0%) | 39.39% (27.6%–51.1%) | 97.85% (94.9%–100%) | 81.16% (76.9%-85.4%) |
| **OR(Mammography AI, Thermalytix AI)** |  | 94.6% (88.6%-100%) | 66.5% (60.9%-72.2%) | 36.6% (28.7%-44.5%) | 98.4% (96.5%-100%) | 80.6% (76.2%-84.9%) |
| **Proposed Density-Informed Multi-modal AI** |  | 94.6% (88.5%–100%) | 79.9% (75.1%–84.7%) | 49.1% (39.5%–58.6%) | 98.6% (97.1%–100%) | **87.2% (83.6%-90.9%)** |

Table II. Performance comparison for unimodal and multi-modal AI systems. (.) indicates 95% CI

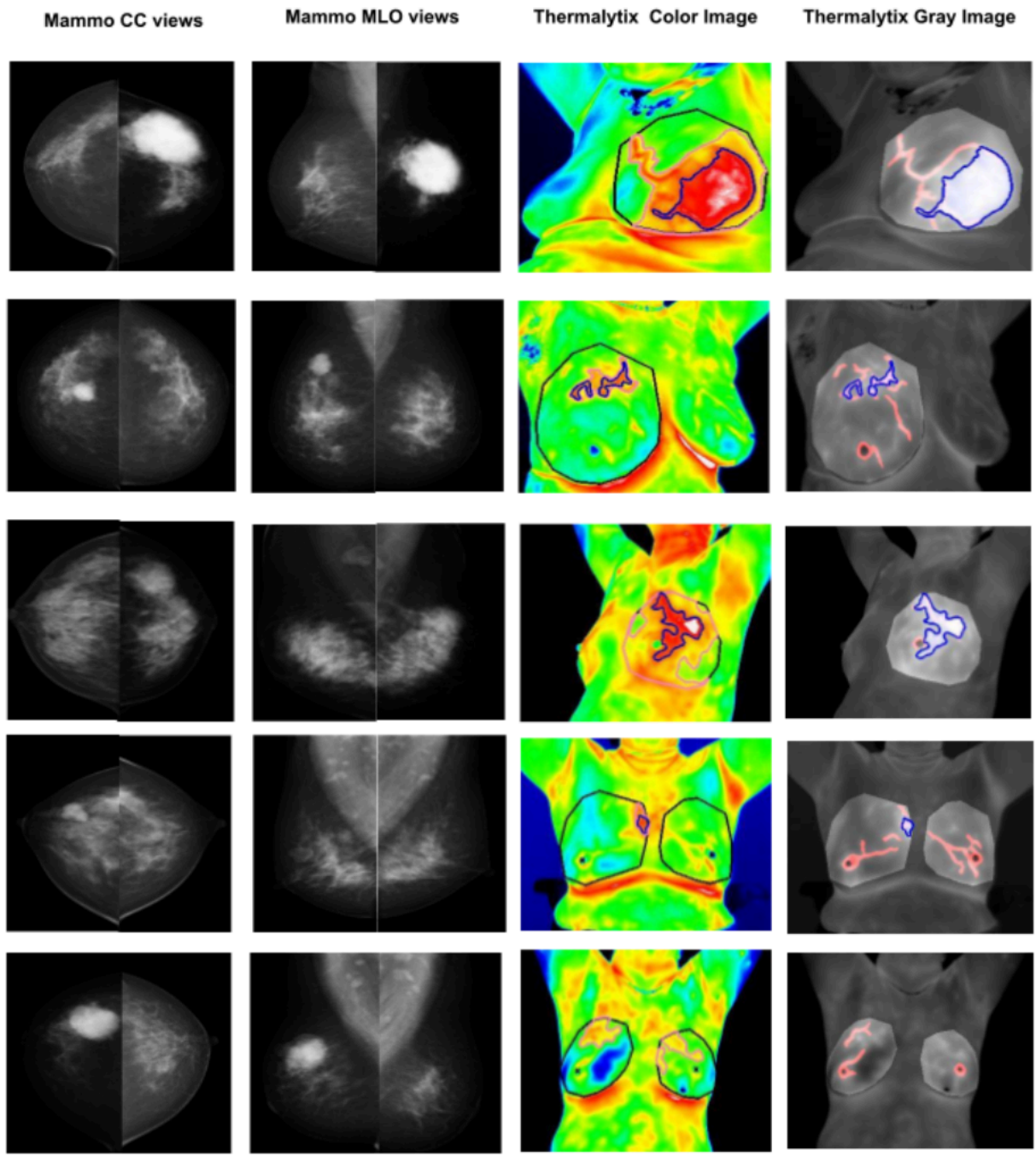

**Figure 7.** Illustration of mammography and corresponding Thermalytix AI images for different participants: (a) Fatty breast tissue: True Positive (TP) for both mammography AI and Thermalytix AI; (b) Dense breast tissue: TP for both mammography AI and Thermalytix AI; (c) Dense breast tissue: TP for Thermalytix AI and False Negative (FN) for mammography AI; (d) Dense breast tissue: TP for Thermalytix AI and FN for mammography AI; (e) Fatty breast tissue: TP for Mammography AI and FN for Thermalytix AI.

## 4. Discussion

Breast cancer screening continues to face critical limitations, particularly in women with dense breast tissue where mammographic sensitivity is markedly reduced. This study introduces and validates a clinically viable, explainable, and scalable multi-modal AI system that addresses these limitations by integrating the anatomical strengths of mammography with the functional insights of thermal imaging through Thermalytix. Our findings reveal that tailoring the imaging modality based on breast density significantly improves diagnostic performance—achieving a sensitivity of 94.55% and a specificity of 79.93% across a diverse screening cohort. This density-informed multi-modal framework closes the long-standing sensitivity gap in dense breasts, outperforming either modality alone and offering a pragmatic path toward more equitable and effective breast cancer detection.

Most current AI efforts in breast cancer screening focus on optimizing performance within a single modality—typically mammography. However, our findings underscore the performance ceiling of unimodal anatomical imaging: while the mammography AI model achieved high sensitivity in fatty breasts (96.3%), its performance dropped to 67.86% in dense tissue, where fibroglandular overlap often masks malignancies. The innovation of our system lies not in marginal algorithmic gains, but in reframing the screening paradigm—by dynamically allocating AI models based on tissue characteristics, we circumvent the intrinsic limitations of single-modality approaches.

This "density-informed AI fusion" represents, to our knowledge, one of the first pragmatic demonstrations of an AI system that adapts to patient-specific anatomy by selecting the most appropriate imaging modality. Routing patients with fatty breasts to mammography AI and those with dense breasts to Thermalytix demonstrates the practical utility of decision-making grounded in breast tissue composition. This rule-based integration is transparent, explainable, and readily implementable in clinical workflows, making it highly attractive for real-world deployment in both high-resource and resource-constrained settings.

Importantly, Thermalytix maintained consistently high sensitivity in both fatty (92.59%) and dense (92.86%) breast types, demonstrating its robustness across tissue categories. This stability is particularly valuable in dense breasts, where anatomical imaging struggles and missed cancers are more likely. These findings reinforce the rationale for assigning Thermalytix AI to dense-breast cases within the hybrid model. While specificity for Thermalytix in dense breasts was lower (69.47%), this requires careful interpretation. A detailed analysis of the 40 false-positive cases revealed that 31 (77.5%) corresponded to benign but clinically relevant findings—such as fibroadenomas, abscesses, and fibrocystic changes. These are not merely algorithmic errors but indications of thermal or vascular abnormalities worthy of clinical attention. In the context of population screening, particularly in high-risk groups, such cautious overcalling may serve as an acceptable and even beneficial trade-off to avoid missed malignancies.

While prior efforts have attempted to combine mammography with ultrasound or MRI to boost sensitivity in dense breasts [39–42], such solutions are often limited by cost, infrastructure, and the need for specialized expertise. In contrast, Thermalytix is affordable, contactless, radiation-free, and operable by minimally trained personnel, making it well-suited as both a primary and adjunctive modality within a multi-modal AI framework.

The public health significance of this multi-modal framework is substantial. Rather than positioning Thermalytix as a standalone alternative, this study demonstrates its value as a complementary modality that enhances screening accuracy when integrated with mammography. In regions with limited access to mammography or radiology expertise, this density-informed multi-modal AI system allows AI-enabled screening to be delivered efficiently and cost-effectively. Even in high-resource settings, its integration could help reduce the burden of interval cancers in women with dense breasts—where up to 40% of malignancies may go undetected by mammography alone.

From a systems perspective, the implementation of this AI-enabled multi-modal pathway introduces minimal workflow complexity while substantially improving diagnostic performance. The decision logic—anchored in a widely available variable (ACR breast density)—is straightforward, automatable, and readily integrable into existing screening infrastructures.

Nonetheless, this study has limitations. First, while the rule-based strategy offers clear interpretability, it may not capture more complex interactions between imaging modalities. Future work should explore joint learning frameworks, including feature-level fusion and end-to-end multi-modal deep learning architectures, to fully exploit cross-modal synergies. Second, validation in larger, more diverse populations across different geographic and clinical contexts is essential to confirm generalizability and support clinical adoption. Additionally, integrating patient risk factors such as age, family history, and hormonal status could further enhance predictive accuracy.

**Conclusion**
In conclusion, this study demonstrates that a density-informed, multi-modal AI framework integrating mammography and Thermalytix significantly improves breast cancer detection across all breast densities. By intelligently assigning imaging modality based on tissue composition, the system addresses long-standing limitations of unimodal screening—particularly in dense breasts—while maintaining high sensitivity, clinical interpretability, and deployment feasibility. This pragmatic and scalable approach offers a compelling path toward more equitable, accurate, and accessible breast cancer screening in both high-resource and resource-limited settings.

**Declaration of generative AI and AI-assisted technologies in the writing process**
During the preparation of this work the author(s) have used ChatGPT in order to correct grammatical errors. After using this tool/service, the author(s) reviewed and edited the content as needed and take(s) full responsibility for the content of publication.